\documentclass[twocolumn,nofootinbib,preprintnumbers,superscriptaddress,amsmath,amssymb,PRL]{revtex4}
\usepackage{graphicx}
\usepackage{dcolumn}
\usepackage{float}
\usepackage{mathptmx, courier, pifont}
\usepackage[scaled=0.92]{helvet}
\usepackage[T1]{fontenc}
\usepackage{textcomp}
\usepackage{color}
\usepackage[normalem]{ulem}

\usepackage[dvipsnames]{xcolor}
\usepackage[colorlinks=true,urlcolor=Blue,linkcolor=Blue]{hyperref}
\usepackage[all]{hypcap}
\usepackage{makecell}

\begin{document}


\title{Observation of an ultra-low $Q$-value electron-capture channel decaying to $^{75}$As via high-precision mass measurement} 
\author{M.~Ramalho}\thanks{Corresponding author: madeoliv@jyu.fi}
\affiliation{Department of Physics, University of Jyv\"askyl\"a, P.O. Box 35, FI-40014  Jyv\"askyl\"a, Finland}%
\author{Z.~Ge}\thanks{Corresponding author. Email address: z.ge@gsi.de}\thanks{Present address: GSI Helmholtzzentrum f\"ur Schwerionenforschung GmbH, 64291 Darmstadt, Germany}
\affiliation{Department of Physics, University of Jyv\"askyl\"a, P.O. Box 35, FI-40014  Jyv\"askyl\"a, Finland}%
\author{T.~Eronen}
\affiliation{Department of Physics, University of Jyv\"askyl\"a, P.O. Box 35, FI-40014  Jyv\"askyl\"a, Finland}%
\author{D.~A.~Nesterenko}
\affiliation{Department of Physics, University of Jyv\"askyl\"a, P.O. Box 35, FI-40014  Jyv\"askyl\"a, Finland}%
\author{J.~Jaatinen}
\affiliation{Department of Physics, University of Jyv\"askyl\"a, P.O. Box 35, FI-40014  Jyv\"askyl\"a, Finland}%
\author{A.~Jokinen} 
\affiliation{Department of Physics, University of Jyv\"askyl\"a, P.O. Box 35, FI-40014  Jyv\"askyl\"a, Finland}%
\author{A.~Kankainen}
\affiliation{Department of Physics, University of Jyv\"askyl\"a, P.O. Box 35, FI-40014  Jyv\"askyl\"a, Finland}%
\author{J.~Kostensalo}
\affiliation{Department of Physics, University of Jyv\"askyl\"a, P.O. Box 35, FI-40014  Jyv\"askyl\"a, Finland}%
\author{J.~Kotila}
\affiliation{Finnish Institute for Educational Research, University of Jyv\"askyl\"a, P.O. Box 35, FI-40014  Jyv\"askyl\"a, Finland}%
\affiliation{Center for Theoretical Physics, Sloane Physics Laboratory Yale University, New Haven, Connecticut 06520-8120, USA}%
\author{M.~I.~Krivoruchenko}
\affiliation{National Research Centre ``Kurchatov Institute'', Ploschad' Akademika
Kurchatova 1, 123182 Moscow, Russia}%
\affiliation{Institute for Theoretical and Experimental Physics, NRC ``Kurchatov
Institute'', B. Cheremushkinskaya 25, 117218 Moscow, Russia}
\author{J.~Suhonen}\thanks{Corresponding author: jouni.t.suhonen@jyu.fi}
\affiliation{Department of Physics, University of Jyv\"askyl\"a, P.O. Box 35, FI-40014  Jyv\"askyl\"a, Finland}%
\author{K.~S.~Tyrin}
\affiliation{National Research Centre ``Kurchatov Institute'', Ploschad' Akademika Kurchatova 1, 123182 Moscow, Russia}%
\author{V.~Virtanen}
\affiliation{Department of Physics, University of Jyv\"askyl\"a, P.O. Box 35, FI-40014  Jyv\"askyl\"a, Finland}%

\date{\today}
\begin{abstract}
A precise determination of the atomic mass of $^{75}$As has been performed utilizing the double Penning trap mass spectrometer, JYFLTRAP. 
The mass excess is measured to be -73035.519(42) keV/c$^2$, which is a factor of 21 more precise and  1.3(9) keV/c$^2$ lower than the adopted value in the newest Atomic Mass Evaluation (AME2020). This value has been used to determine the ground-state-to-ground-state electron-capture decay $Q$ value of $^{75}$Se and $\beta^-$ decay $Q$ value of $^{75}$Ge, which 
are derived to be 866.041(81) keV and 1178.561(65) keV, respectively. 
Using the nuclear energy-level data of 860.00(40) keV, 865.40(50) keV (final states of electron capture) and 1172.00(60) keV (final state of $\beta^-$ decay) for the excited states of $^{75}$As$^*$, we have determined the ground-state-to-excited-state $Q$ values for two transitions of $^{75}$Se  $\rightarrow$ $^{75}$As$^*$ and one transition  of $^{75}$Ge $\rightarrow$ $^{75}$As$^*$.
 The ground-state-to-excited-state $Q$ values are determined to be 6.04(41) keV, 0.64(51) keV and 6.56(60) keV, respectively, thus confirming that the three low $Q$-value transitions are all energetically valid and one of them is a possible candidate channel for antineutrino mass determination.
 Furthermore, the ground-state-to-excited-state $Q$ value of transition $^{75}$Se $\rightarrow$ $^{75}$As$^*$ (865.40(50) keV) is revealed to be ultra-low (< 1 keV) and the first-ever confirmed EC transition possessing an ultra-low $Q$ value from direct measurements.
\end{abstract}
\maketitle
\section{Introduction}

The discovery of flavor oscillations of atmospheric, solar, and reactor neutrinos confirms that neutrinos have mass~\cite{Fukuda1998,SNOCollaboration2002,Gerbino2018a}. The standard model (SM) contradictorily predicts that the neutrino mass is zero. How neutrinos acquire their small masses is consequently a matter of great theoretical interest and may be an evidence of new physics beyond the SM. 
Oscillation data provide only the differences of the squared neutrino masses but not their absolute values.
In order to solve the open problem of the absolute values of the neutrino masses, laboratory measurements are necessary.
Among these experiments, the ones dedicated to measurements of the neutrinoless double beta decay aim to determine if neutrinos are of Dirac or Majorana nature and to measure the effective Majorana neutrino mass~\cite{Suhonen1998,Avignone2008,Ejiri2019,Blaum2020}.  Unfortunately, this method is model-dependent and highly relies on the calculated values of the involved transition matrix elements.
A direct and model-independent method for determining (anti)neutrino mass is based on single beta decay or electron capture (EC)~\cite{Gastaldo2017,aker2021direct}. 
From the $\beta^-$ decay of tritium, one can determine the electron-antineutrino mass by zooming in the slight distortion of the shape of the electron spectrum near the endpoint, determined by the decay $Q$ value minus the antineutrino mass.
In a similar vein, in the EC of $^{163}$Ho \cite{Gastaldo2017}, one can determine the electron-neutrino mass from the endpoint of the measured de-excitation spectrum, which is shifted below $Q$ by the non-zero neutrino mass. In both beta decay and EC, the sensitivity to (anti)neutrino mass is increased  by a small $Q$ value. 
Therefore, as small as possible $Q$ value is desired in these (anti)neutrino-mass determination experiments. Due to this reason, tritium with low $\beta^-$-decay $Q$ value of 18.59201(7) keV~\cite{Myers2015} in the KATRIN (KArlsruhe TRitium Neutrino) experiment~\cite{Drexlin2013,Aker2019}, and $^{163}$Ho with low EC $Q$ value of  2.833(30)$_{sta}$(15)$_{sys}$ keV~\cite{Eliseev2015} in the ECHo (Electron Capture in $^{163}$Ho)~\cite{Gastaldo2014,Gastaldo2017} HOLMES~\cite{Alpert2015,Faverzani2016} experiments are used.
An upper limit of 0.8 eV/c$^2$ (90\% Confidence Level (C.L.)) for the electron-antineutrino mass  is achieved in the tritium decay by KATRIN~\cite{aker2021direct}, and  for electron-neutrino mass an upper limit of 150 eV/c$^2$ (95\% C.L.) was obtained exploiting the EC of $^{163}$Ho in the ECHo experiment~\cite{Velte2019}. 

Other isotopes with low $Q$-value decay transition, especially ultra-low (< 1 keV), are of significant interest for  future neutrino-mass absolute-scale determination experiments~\cite{Mustonen2010,Mustonen2011,Suhonen2014,deRoubin2020,ge2021,ge2021b}.  The  $Q$ values determined with indirect methods may include unknown systematic uncertainties. This can make these $Q$ values deviate by more than 10 keV~\cite{ge2021,Fink2012,Nesterenko2019} from those directly determined through high-precision Penning-trap mass spectrometry (PTMS). Direct measurements of the masses or $Q$ values  through PTMS, which is to date the only direct method to have achieved $\sim$ 100 eV precision or better, are indispensable in the searches for ultra-low $Q$-value transitions.
The first ultra-low $Q$-value decay, $^{115}$In (9/2$^{+}$, ground state)  $\rightarrow$ $^{115}$Sn (9/2$^{+}$, excited state), was discovered by Cattadori \emph{et al.}~\cite{Cattadori2005} and the  ultra-low $Q$ value was confirmed by the JYFLTRAP Penning trap ~\cite{Wieslander2009} and the FSU Penning trap~\cite{Mount2009}. 
The second ultra-low $Q$-value case of $^{135}$Cs (7/2$^{+}$, ground state)  decaying to $^{135}$Ba (11/2$^{-}$, second excited state)   was discovered at JYFLTRAP~\cite{deRoubin2020,Kankainen2020} recently.   
These are the only two cases that are confirmed to possess ultra-low $Q$ values from direct measurements, and they both belong to the category of $\beta^{-}$ decay.

In this paper, we report on the  high-precision mass determination of  $^{75}$As from cyclotron frequency measurements with high-precision PTMS.  The ground-state-to-ground-state (gs-to-gs) EC $Q$ value of the transition $^{75}$Se  $\to\,^{75}$As and $\beta^{-}$-decay $Q$ value of the transition $^{75}$Ge $\to\,^{75}$As are determined.  The high-precision gs-to-gs $Q$ values  from this work, combined with the nuclear energy-level data  for three excited states of $^{75}$As, are used to determine their ground-state-to-excited-state (gs-to-es) $Q$ values. Three low $Q$-value transitions are confirmed to be energetically possible and  the first to date ultra-low $Q$-value EC transition has been discovered in this experiment.

\begin{figure*}[!htb]
\centering
\includegraphics[width=1.99\columnwidth]{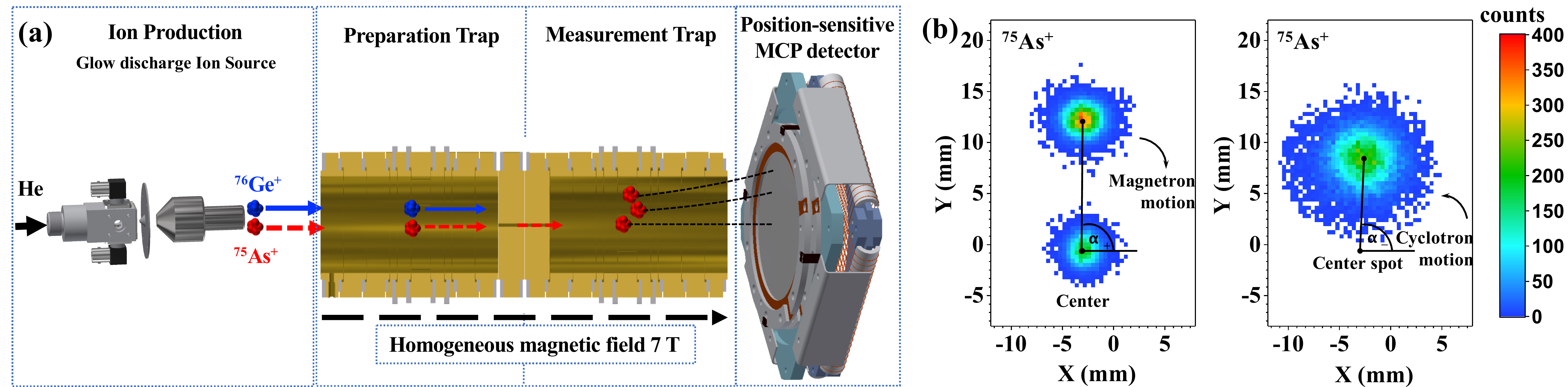}
\caption{(Color online).  (a) Schematic view of ion production and mass measurements with PI-ICR technique. The stable $^{75}$As$^+$ and $^{76}$Ge$^+$ ions were simultaneously produced with an offline glow-discharge ion source, where the ions were produced and transported with an He gas flow and electric fields. 
Ions having mass number of 75 or 76 were selected with a dipole magnet  and transported to the JYFLTRAP PTMS for final ion species selection in the  preparation trap by means of a buffer-gas cooling and cyclotron frequency determination using the phase-imaging technique at the measurement trap. A position-sensitive MCP detector was used to register the images of the phases. 
(b) An illustration of the radial-motion ("magnetron", "cyclotron", and "center") projection of the $^{75}$As$^+$ ions onto the position-sensitive MCP detector. 
The magnetron phase spot is displayed on the left side and the cyclotron phase spot on the right. The angle difference between the two spots relative to the center spot is utilized to deduce the cyclotron frequency of the measured ion. The number of ions in each pixel is illustrated by color bars.
} 
\label{fig:igisol}
\end{figure*}

\section{Experimental method}
The atomic mass of $^{75}$As  was measured with the JYFLTRAP double PTMS~\cite{Eronen2012} at the Ion Guide Isotope Separator On-Line facility (IGISOL)~\cite{Moore2013} of the University of Jyv\"askyl\"a. 
The stable $^{75}$As$^+$ ions were produced with an offline glow-discharge ion source. To measure the mass of $^{75}$As precisely, $^{76}$Ge$^+$ ions with well-known mass value were used as reference ion, co-produced from the same ion source.
As shown in Fig.~\ref{fig:igisol} (a), the gas cell of the glow-discharge ion source contains  two sharp electrodes,  of which one is made of the mixture of natural abundance arsenic and germanium, allowing simultaneous production of ions of these elements.
The produced ions were stopped in the gas cell and extracted out with gas flow and electric fields. 
They were subsequently accelerated with a voltage of $\sim$ 30~kV and transported further to a 90$^\circ$ electrostatic bender to the main horizontal beamline before the dipole magnet.
With a mass resolving power of  $M/\Delta{M}$ $\sim$ 500, the  dipole magnet is used to selectively transport the  $^{75}$As$^+$ or  $^{76}$Ge$^+$ ions to a radiofrequency quadrupole cooler-buncher (RFQ)~\cite{Nieminen2001}.
After cooling and bunching the ions with the RFQ, they were transported to the JYFLTRAP double PTMS as schematically shown in Fig.~\ref{fig:igisol} (a).

After the dipole magnet, in principle pure ion beams of $^{75}$As$^+$ and $^{76}$Ge$^+$ with unambiguous identification from the offline ion source are separated and selected. To ensure that any possible isobaric or molecular contaminations are removed, the first (preparation) trap is used via the sideband buffer gas cooling technique~\cite{Savard1991} with a typical resolving power of around $10^{5}$. Finally, the purified sample of $^{75}$As$^+$ or $^{76}$Ge$^+$ ions are  injected into the second (measurement) trap.
The ion's cyclotron frequency 
$\nu_{c}=\frac{1}{2\pi}\frac{qB}{m}$,
where $B$ is the magnetic field strength, $q$ is the charge state and $m$ the mass of the ion, is determined in the second Penning trap.


The phase-imaging ion-cyclotron-resonance (PI-ICR) technique~\cite{Nesterenko2018} at JYFLTRAP is used to measure the cyclotron frequency in this work. 
This technique depends on projecting the ion motion in the Penning trap onto a position-sensitive multichannel-plate ion detector (schematic shown in Fig.~\ref{fig:igisol} (a)) and provides around 40 times better resolving power and is 25 times faster than the conventional TOF-ICR method~\cite{Nesterenko2018,Eliseev2014,Eliseev2013}. Measurement scheme 2 described in~\cite{Eliseev2014} was applied to measure the cyclotron frequency $\nu_{c}$ of the corresponding nuclide. 

The PI-ICR measurement is initiated by  exciting a cooled ion bunch with a dipolar radio-frequency electric field at the trap-modified cyclotron frequency $\nu_{+}$. This is a short excitation lasting approximately 1~ms. Afterwards, the trap-modified cyclotron motion  is converted to magnetron motion using a quadrupole excitation with the cyclotron frequency $\nu_c$ and a duration of approximately 2~ms. Two different delays for the application of this excitation are alternatively used. 
In one, the conversion pulse is applied right after the dipolar excitation and in the other after a longer time. The time difference of the excitation pulses is called the phase accumulation time, denoted $t$. 
During this time, in a cycle with a very short conversion pulse delay, the ions perform magnetron motion and in the one with a long delay modified cyclotron motion.  
After, the ions are extracted from the trap (the time difference between the initial dipolar excitation and the extraction is always constant irrespective of the quadrupole conversion delay) and their positions are recorded with a position-sensitive MCP detector~\cite{PS-MCP}. The spot obtained with the small conversion time delay is called the magnetron spot and the one with the long delay is the cyclotron spot. 
The center spot  of  the ions of interest is collected by extracting the ions directly to project them onto the MCP detector without applying any excitation. 
The angle between the spots of the cyclotron and magnetron phases with respect to the center spot is $\alpha_c$ = $\alpha_+$-$\alpha_-$, where $\alpha_+$ and $\alpha_-$ are the polar angles of  the magnetron motion phase and cyclotron motion phases, respectively. The cyclotron frequency $\nu_{c}$  is deduced from: 
\begin{equation}
\label{eq:nuc2}
\nu_{c}=\frac{\alpha_{c}+2\pi n_{c}}{2\pi{t}},
\end{equation}
where $n_{c}$ are the number of complete revolutions of the measured ion during the phase accumulation time $t$. The accumulation time $t$ for $^{75}$As$^{+}$ and  $^{76}$Ge$^{+}$ was 400 ms, which ensures the spot of interest was resolved from any possible isobaric, and molecular contamination. The positions of the magnetron-motion and cyclotron-motion phase spots were chosen such that the angle $\alpha_c$  did not exceed a few degrees.  By this way, the shift in the frequency ratio measurements, due to the conversion of the cyclotron motion to magnetron motion and the possible distortion of the ion-motion projection, can be reduced to a level well below 10$^{-10}$~\cite{Eliseev2014} . 
In addition, the start of the initial dipolar excitation with frequency $\nu_{+}$ was repeatedly scanned over one magnetron period (6 points) and the extraction was scanned over one cyclotron period (6 points) to average out any residual magnetron and cyclotron motion that could shift the different spots. 
One collected data of phase imaging of  on the MCP detector on Fig.~\ref{fig:igisol} (b).

The atomic mass of the ion of interest was derived from the measured cyclotron frequency ratio:
\begin{equation}
\label{eq:mass}
M_{ioi} = R(M_{ref} - m_e)   + m_e  + (R \cdot B_{ref} - B_{ioi})/c^2,
\end{equation}
where $M_{ioi}$ and  $M_{ref}$ are the masses of the ion of interest  ($^{75}$As$^{+}$) and reference  ($^{76}$Ge$^{+}$) atoms, respectively.  $R = \frac{\nu_{c,{ref}}}{\nu_{c,ioi}}$ is their cyclotron frequency ratio for singly charged ions. $m_{e}$ is the mass of electron and $c$ is the speed of light in vacuum. $B_{ref}$ and $B_{ioi}$ are the electron binding energy of $^{76}$Ge$^{+}$ and $^{75}$As$^{+}$, which are 7.899440(10) eV and 9.78855(25) eV from~\cite{NIST_ASD}, respectively. The main contributions of the final mass uncertainty of $^{75}$As are from the statistical uncertainty of the measurements of $R$ and the reference mass uncertainty (0.018 keV from~\cite{Wang2021}).

\begin{table}[!htb]
\caption{Final results from the analysis of mean cyclotron frequency ratio between the ion of interest and reference nuclei.  The first column gives lists of ion of interest and the reference (IOI-Ref) nuclei.  The measured frequency ratio $\overline{R}$ is in the second column. The mass excess (ME) in keV/c$^2$ of ion of interest determined in this work in comparison to the AME2020 values~\cite{Wang2021} are listed in the third column and splited in two rows. The last column demonstrates the difference of the ME value from this work and that adopted from AME2020.}
\begin{ruledtabular}
   \begin{tabular*}{0.48\textwidth}{@{}ccccc@{}}
IOI-Ref&$\overline{R}$  & \makecell[c]{ME (this work) \\ME (AME2020)}&Diff. \\
\hline\noalign{\smallskip}
$^{75}$As-$^{76}$Ge&  0.986 830 896 52(53) & \makecell[c]{-73035.521(42)\\-73034.20(90) }& -1.32(90)\\
\hline
$^{77}$Se-$^{76}$Se&  1.013 181 218 27(79) & \makecell[c]{-74599.443(58)  \\-74599.490(60)} &0.047(84) \\
\hline
$^{94}$Mo-$^{95}$Mo&  0.989 455 235 86(72) & \makecell[c]{-88414.101(136)\\-88414.06(14)} &-0.041(195)\\
   \end{tabular*}
   \label{table:ME}
\end{ruledtabular}
\end{table}

\begin{table*}[!htb]
   \caption{Transitions from the ground states of parent nuclei $^{75}$Ge and $^{75}$Se to the excited states of the daughter $^{75}$As. 
 The first  and second columns illustrate the experimental spin-parities of the initial ground states and the half-lives of the parent nuclei.
The third column gives the measured spin-parities of the excited final states for the transitions. The fourth  column gives the decay type.  The fifth column gives the gs-to-gs decay $Q$ values ($Q_{\beta^-}^0$ for $\beta^-$-decay and $Q_{EC}^0$ for EC) from literature (AME2020)~\cite{Wang2021} and the sixth column from this work. The seventh and eighth columns give the gs-to-es decay $Q$ values ($Q_{}^*$) from literature and this work, respectively. The  last  column gives the excitation energies E$^{*}$ from~\cite{NNDC}. All the energy values are in units of keV. Spin-parity assignments enclosed in braces indicate that these are uncertain, which results in an uncertainty in the decay type, indicated by a  \{?\}. FNU denotes forbidden non-unique.
 }
  \begin{ruledtabular}
   \begin{tabular*}{\textwidth}{@{}ccccccccc@{}}
Initial state & Half-life&Final state &Decay type & \makecell[c]{$Q_{}^0$ \\(AME2020)} &\makecell[c]{$Q_{}^0$ \\ (This work)}& \makecell[c]{$Q_{}^*$ \\(AME2020)} &\makecell[c]{$Q_{}^*$ \\ (This work)}& $E^{*}$ \\
\hline\noalign{\smallskip} 
       $^{75}$Se (5/2$^{+}$)&119.78(5) d&   $^{75}$As (\{3/2$^{-}$, 5/2$^{-}$\})& EC: 1st FNU & 864.72(90) &866.041(81) &   -0.7(10) &0.64(51)&865.40(50)\\
        $^{75}$Se (5/2$^{+}$)&119.78(5) d&   $^{75}$As (1/2$^{+}$)& EC: 2nd FNU& 864.72(90) &866.041(81) &   4.7(10) &6.04(41)&860.00(40)\\
\hline
    $^{75}$Ge (1/2$^{-}$) &82.78(4) m& $^{75}$As (\{1/2$^{-}$:7/2$^{-}$\})&  $\beta^{-}$:  Allowed\{?\} &    1177.24(90)&1178.561(65)& 5.2(11) &6.56(60)   &1172.00(60) \\
   \end{tabular*}
   \label{table:low-Q}
   \end{ruledtabular}
\end{table*}

\begin{figure}[!htb]
   \includegraphics[width=0.99\columnwidth]{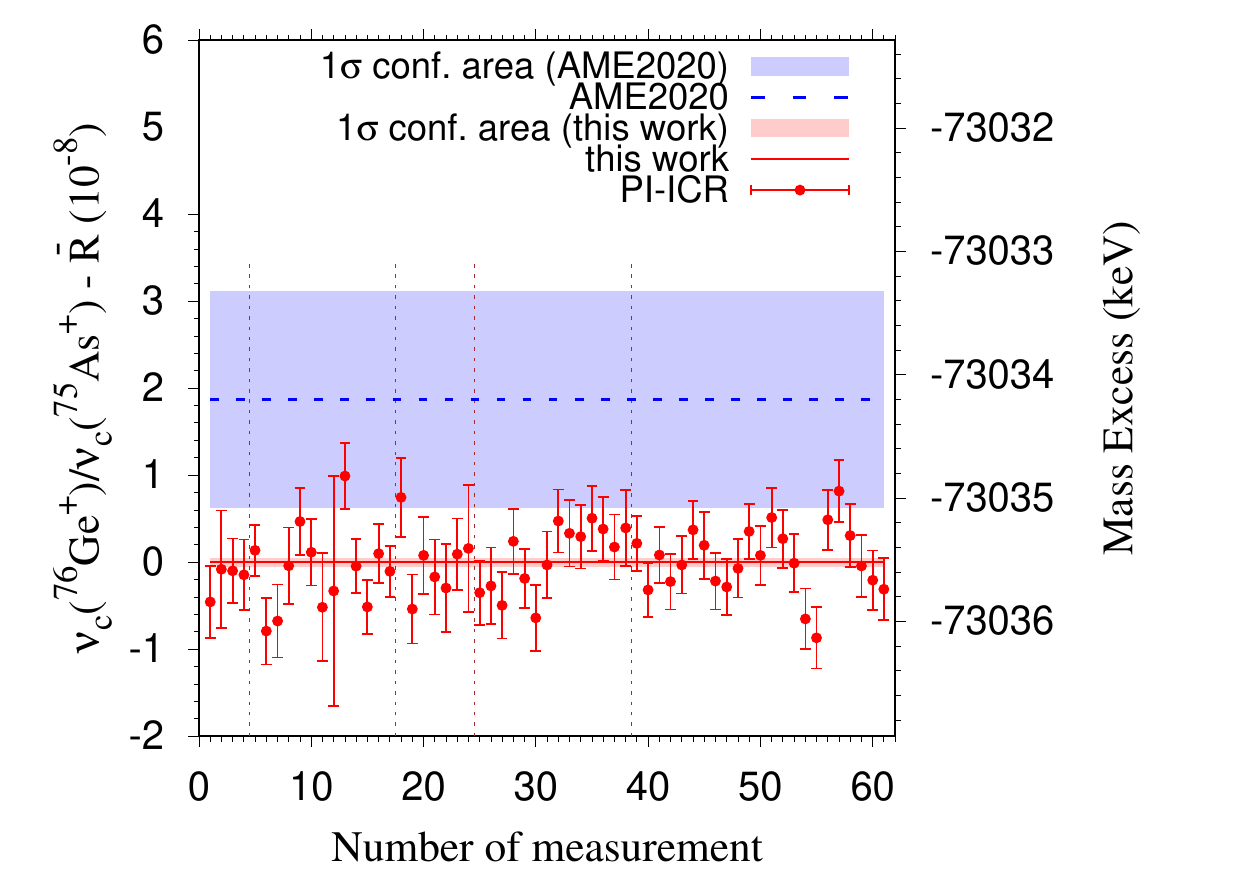}
   \caption{(Color online). The measured cyclotron frequency ratios  $R$ ( $\nu_c$($^{76}$Ge$^{+}$)/$\nu_c$($^{75}$As$^{+}$) (left axis) and mass excess (right axis) in this work compared to values adopted from AME2020. The red dots with uncertainties are the measured PI-ICR single ratios in five time slots, which are separated with vertical brown dashed lines. The weighted average value in this work  $\overline{R}$ = 0.986 830 896 52(53) represents by the solid red line and its 1$\sigma$ uncertainty band is shaded in red. The dashed blue line indicates the value adopted from AME2020  with its 1$\sigma$  uncertainty area shaded in blue.}
   \label{fig:ratio}
\end{figure}

\section{Results and discussion}
\subsection{$Q$ value determination}
Five data sets for PI-ICR measurements were performed in different time slots. Every data set was accumulated by switching between the ion species $^{75}$As$^{+}$ and  $^{76}$Ge$^{+}$ every four cycles.
Each cycle took less than 2 minutes to complete a full scanning measurements of the magnetron phase, cyclotron phase and center spot in sequence. In the analysis, typically 8 to 16 cycles were summed before determining the position of each spot by fitting with the maximum likelihood method. 
The phase angles were calculated accordingly to deduce the cyclotron frequencies of each ion species. Closest measured cyclotron frequencies $\nu_{c}$ of the reference  was  linearly interpolated to the time for measurement of the ion of interest to deduce the cyclotron frequency ratio $R$. Only events less than five ions/bunch  were considered in the analysis  in order to reduce the possible cyclotron frequency shifts due to ion-ion interaction~\cite{Kellerbauer2003,Roux2013}.
A countrate-class analysis~\cite{Roux2013} was performed and the count rate related frequency shifts were not observed in the analysis. A temporal fluctuation of the magnetic field $\delta_B(\nu_{ref})/\nu_{ref}=  \Delta t \times 2.01(25)  \times  10^{-12}$/min~\cite{nesterenko2021}, where $\Delta t$ is the time interval between two consecutive reference measurements, is considered in the analysis. Furthermore,  a mass-dependent uncertainty of $\delta_m(r)/r=\Delta m \times2.35(81)\times10^{-10}/u$~\cite{nesterenko2021}, was taken into account as there is one mass unit (1 $u$) difference of the two measured ion species. To check whether some other systematic error should be added due to this technique, a crosscheck~\cite{Nesterenko2020} has been carried out using well-known mass pairs $^{77}$Se-$^{76}$Se and $^{94}$Mo-$^{95}$Mo~\cite{Huang2021,Wang2021}, which both have one mass unit difference. The measured results are shown in Table~\ref{table:ME}. The measured mass excesses of both cases are consistent with AME2020 values within 1$\sigma$ uncertainty. Therefore, in our final results, we did not include any further systematic error.

The weighted mean ratio $\overline{R}$ of all single ratios was calculated along with the inner and outer errors to deduce the Birge rato~\cite{Birge1932}.  The maximum of the inner and outer errors was taken as the weight to calculate $\overline{R}$.  
Results of the analysis including all PI-ICR data sets in comparison to literature values are demonstrated in Fig.~\ref{fig:ratio}. The final frequency ratio $\overline{R}$ as well as the derived mass-excess value are respectively 0.986 830 896 52(53) and  -73035.521(42)  keV/c$^2$. 

\begin{figure}[!htb]
   \includegraphics[width=1.0\columnwidth]{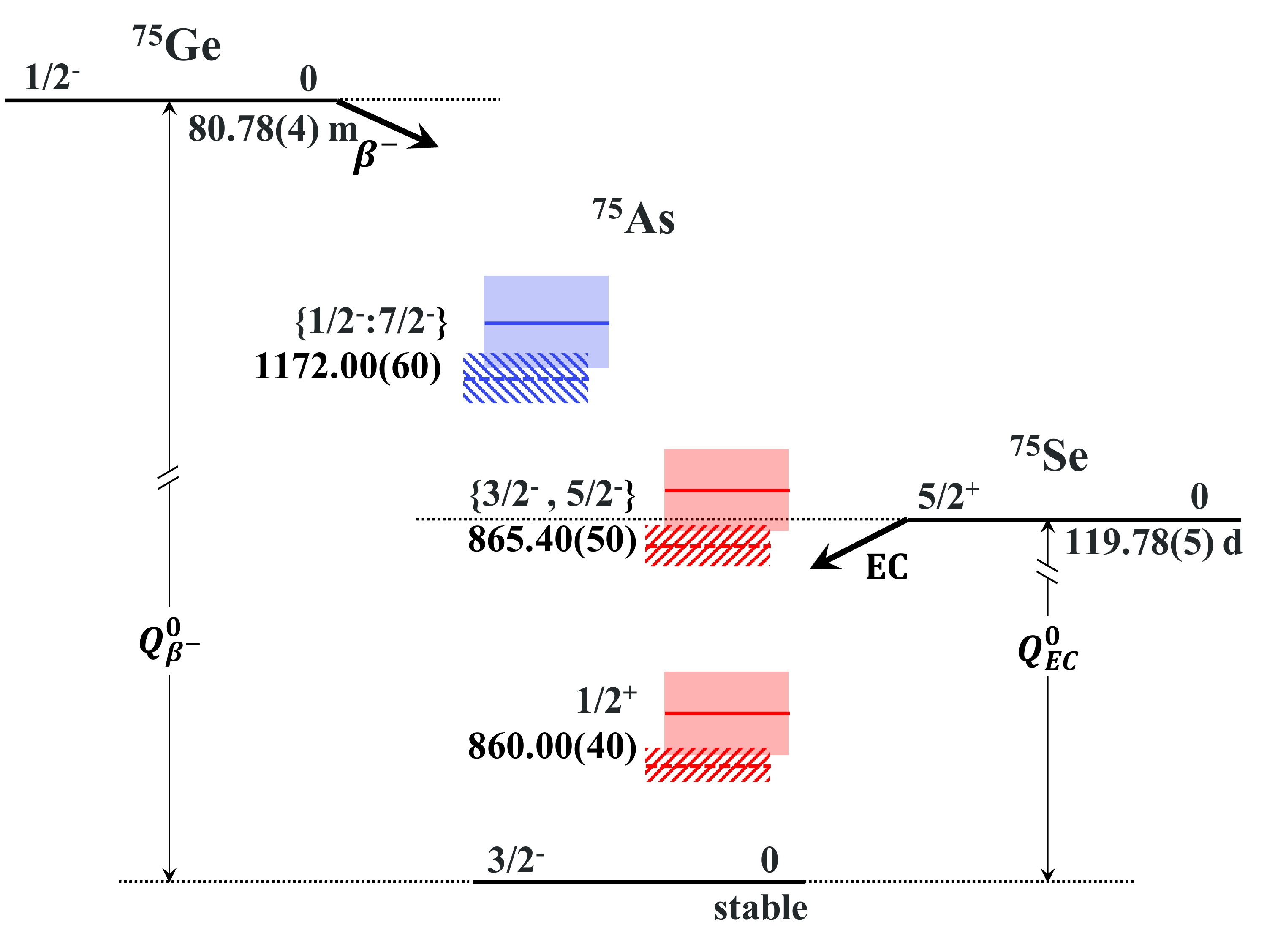}
   \caption{(Color online). 
   Partial decay scheme of $^{75}$Se and $^{75}$Ge to $^{75}$As using $Q$ values from AME2020~\cite{Huang2021,Wang2021} in comparison to this work. The energies of the excited states and spin-parites indicated in the figure are from~\cite{NNDC}. The solid lines demonstrate the levels of the excited states with the $Q$ values from AME2020 and dashed lines from the refined  $Q$ values in this work.
 The shaded or hatched areas (in blue for $\beta^-$ decay or in red for EC) illustrate the corresponding 1 $\sigma$ uncertainties of the $Q$ values. 
 See  Table.~\ref{table:low-Q} for more details on the $Q$ values.
 }
   \label{fig:level-scheme}
\end{figure}
%
%

The mass-excess value of $^{75}$As  in AME2020 is evaluated from two indirect reaction measurements, $^{75}$As(p,n)$^{75}$Se and $^{78}$Se(p,a)$^{75}$As, which have the influences of 85.3\% and 14.7\% on the primary nuclide, respectively.
The refined mass-excess value of  $^{75}$As from our direct measurements is 21 times more precise. The value is 1.32(90) keV/c$^2$ lower than that adopted in AME2020 and indicates that $^{75}$As is more bound. 
The high-precision mass-excess value of  $^{75}$As was used to deduce the gs-to-gs $Q$   values ($Q_{}^0$ deduced from the mass difference of the decay-parent and decay-daughter nuclei) of the $^{75}$Se $\rightarrow$ $^{75}$As and $^{75}$Ge $\rightarrow$ $^{75}$As transitions.
Combined with the excitation energies $E^*$ from~\cite{NNDC} for the excited states of $^{75}$As, the gs-to-es $Q$ values ($Q_{}^*$ = $Q_{}^0$- $E^*$) were calculated. 
The comparison of the $Q$ values from this work to the corresponding $Q$ values from literature is performed in Table.~\ref{table:low-Q}. 
The presently measured gs-to-es $Q$ values of the transitions tabulated in Table~\ref{table:low-Q} are determined with  uncertainties less than 100 eV, while the uncertainties of the gs-to-es $Q$ values  are much larger than 100 eV, the main contributions coming from the large uncertainties of the excitation energies of the final states. 
Our results confirm that three low $Q$-value transitions from the ground-states of $^{75}$Se and $^{75}$Ge to the corresponding excited states of $^{75}$Se, as tabulated in Table.~\ref{table:low-Q},  are all energetically allowed.
Furthermore, our measurements confirm that the $Q_{}^*$ value corresponding to the EC transition $^{75}$Se $\rightarrow$ $^{75}$As* (865.40(50) keV) is ultra-low and positive at a level of $\sim$ 1.3$\sigma$. This transition and the transition to the $1/2^+$ state at 860.00(40) keV are non-unique EC transitions and thus their characteristics depend on nuclear-structure details through the nuclear matrix elements \cite{Ejiri2019,Behrens1982}. This excludes their use as reliable sources of information about the electron-neutrino mass. On the other hand, the $\beta^-$ decay to the excited state at 1172.00(60) keV might be of interest for the electron-antineutrino mass measurements if the spin of the state is either $1/2^-$ or $3/2^-$, corresponding to an allowed $\beta^-$ transition. Hence, accurate determination of the excitation energy and spin of the state is highly important and called for.

\subsection{Beta-decay study via nuclear shell-model}
The low-$Q$ EC transitions of interest go to a $1/2^+$ state at 860.00(40) keV and a negative-parity state at 865.40(50) keV, experimentally either a $3/2^-$ or a $5/2^-$ state. The $\beta^-$ transition goes to a negative-parity state at 1172.00(60) keV, the experimental spin assignment being between a $1/2^-$ and a $7/2^-$ state. One may try to shed light on these spin-assignment ambiguities by performing
nuclear shell-model (NSM) calculations. States of the isotope $^{75}$As were computed using the software $\textit{NuShellX}$~\cite{Brown2014} in a single-particle model space consisting of the orbitals $1f_{5/2}$, $2p_{3/2}$, $2p_{1/2}$, and $1g_{9/2}$ for both neutrons and protons. Effective single-particle energies have been fitted to be used within the region of interest $A$ = 63 - 96 in~\cite{Honma2009}. In the present study both jun45pn and jj44bpn interactions were applied. The interaction jun45pn was originally applied to a neighbouring isotope ($^{76}$Ge)~\cite{Mukhopadhyay2017} while the jj44bpn was fitted to work within the presently studied region~\cite{Honma2009}.

The sequence of the spin-parities of the levels in $^{75}$As are well reproduced by both interactions up to the experimental negative-parity state at 865.40(50) keV.
The level energies predicted by the NSM agree with the experimental ones within a range of 50 to 200 keV. The two interactions suggest consistently the spin assignment $5/2^-$ for the experimental 865.40(50) keV state. The corresponding EC transition is thus predicted to be a first-forbidden non-unique $\Delta J=0$ transition and thus its properties depend sensitively on nuclear-structure details.

Concerning the experimental state at $1172.00$ keV, jun45pn predicts a
$3/2^{-}$ state at $910.0$ keV and a $5/2^{-}$ state at $1296.0 $ keV. 
This suggests that the most likely spin-assignments for the $1172.0$ keV state are either $3/2^{-}$ or $5/2^{-}$.
The jj44bpn interaction predicts a $7/2^{-}$ state at $1129.0 $ keV and a $3/2^{-}$ state at $1269.0 $ keV. 
Considering the predictions of both interactions together leaves the $3/2^{-}$ as the most likely spin-parity for the $1172.00$ keV state. This, in turn, makes the $\beta^-$ transition from the $1/2^-$ state of $^{75}$Ge to this state an allowed one,
with an available $Q_{\beta^-}$ of 6.56(60) keV. This would make the transition interesting from the point of view of electron-antineutrino mass measurements. Eventually, a re-measurement of the spin-parities of the states in question is called for. 

\section{Conclusion}
The atomic mass of $^{75}$As is precisely measured directly with the PI-ICR technique at the JYFLTRAP double PTMS. The precision of the $^{75}$As mass has been improved by a factor of 21 and was found to be 1.32(90) keV/c$^2$  more bound than the evaluated value in the AME2020 mass evaluation. This allows a high-precision determination of the gs-to-gs $Q$ values of the decays of $^{75}$Ge and $^{75}$Se to $^{75}$As with an uncertainty of better than 100 eV. Exploiting the energy-level data of the excites states of $^{75}$As, the gs-to-es $Q$ values of three low $Q$-value transitions have been determined to a sub-keV precision. The derived positive gs-to-es $Q$ values have all been confirmed to be associated with energetically-allowed low $Q$-value EC and $\beta^-$ transitions. Among them, a first ultra-low (< 1 keV) $Q$-value EC transition has been found. One of the transitions, namely the low-$Q$ $\beta^-$ transition, has the potential to be used in determination of the value of the electron-antineutrino mass. For this to be realized, a more precise determination of the excitation energy and spin of the corresponding final state is of paramount importance.

\acknowledgments 
We acknowledge the support by the Academy of Finland under the Finnish Centre of Excellence Programme 2012-2017 (Nuclear and Accelerator Based Physics Research at JYFL) and projects No. 306980, 312544, 275389, 284516, 295207, 314733, 318043, 327629 and 320062. The support by the EU Horizon 2020 research and innovation program under grant No. 771036 (ERC CoG MAIDEN) is acknowledged.

\bibliography{my-final-bib-from-jabref}

\begin{thebibliography}{50}
\expandafter\ifx\csname natexlab\endcsname\relax\def\natexlab#1{#1}\fi
\expandafter\ifx\csname bibnamefont\endcsname\relax
  \def\bibnamefont#1{#1}\fi
\expandafter\ifx\csname bibfnamefont\endcsname\relax
  \def\bibfnamefont#1{#1}\fi
\expandafter\ifx\csname citenamefont\endcsname\relax
  \def\citenamefont#1{#1}\fi
\expandafter\ifx\csname url\endcsname\relax
  \def\url#1{\texttt{#1}}\fi
\expandafter\ifx\csname urlprefix\endcsname\relax\def\urlprefix{URL }\fi
\providecommand{\bibinfo}[2]{#2}
\providecommand{\eprint}[2][]{\url{#2}}

\bibitem[{\citenamefont{Fukuda et~al.}(1998)\citenamefont{Fukuda, Hayakawa,
  Ichihara, Inoue, Ishihara, Ishino, Itow, Kajita, Kameda, Kasuga
  et~al.}}]{Fukuda1998}
\bibinfo{author}{\bibfnamefont{Y.}~\bibnamefont{Fukuda}},
  \bibinfo{author}{\bibfnamefont{T.}~\bibnamefont{Hayakawa}},
  \bibinfo{author}{\bibfnamefont{E.}~\bibnamefont{Ichihara}},
  \bibinfo{author}{\bibfnamefont{K.}~\bibnamefont{Inoue}},
  \bibinfo{author}{\bibfnamefont{K.}~\bibnamefont{Ishihara}},
  \bibinfo{author}{\bibfnamefont{H.}~\bibnamefont{Ishino}},
  \bibinfo{author}{\bibfnamefont{Y.}~\bibnamefont{Itow}},
  \bibinfo{author}{\bibfnamefont{T.}~\bibnamefont{Kajita}},
  \bibinfo{author}{\bibfnamefont{J.}~\bibnamefont{Kameda}},
  \bibinfo{author}{\bibfnamefont{S.}~\bibnamefont{Kasuga}},
  \bibnamefont{et~al.}, \bibinfo{journal}{Physical Review Letters}
  \textbf{\bibinfo{volume}{81}}, \bibinfo{pages}{1562} (\bibinfo{year}{1998}),
  ISSN \bibinfo{issn}{10797114}, \eprint{9807003},
  \urlprefix\url{http://link.aps.org/doi/10.1103/PhysRevLett.81.1562}.

\bibitem[{\citenamefont{{SNO Collaboration}}(2002)}]{SNOCollaboration2002}
\bibinfo{author}{\bibnamefont{{SNO Collaboration}}}, \bibinfo{journal}{Physical
  Review Letters} \textbf{\bibinfo{volume}{89}}, \bibinfo{pages}{1}
  (\bibinfo{year}{2002}), ISSN \bibinfo{issn}{10797114}, \eprint{0204008},
  \urlprefix\url{http://dx.doi.org/10.1103/PhysRevLett.89.011301}.

\bibitem[{\citenamefont{Gerbino and Lattanzi}(2018)}]{Gerbino2018a}
\bibinfo{author}{\bibfnamefont{M.}~\bibnamefont{Gerbino}} \bibnamefont{and}
  \bibinfo{author}{\bibfnamefont{M.}~\bibnamefont{Lattanzi}},
  \bibinfo{journal}{Frontiers in Physics} \textbf{\bibinfo{volume}{5}}
  (\bibinfo{year}{2018}).

\bibitem[{\citenamefont{Suhonen and Civitarese}(1998)}]{Suhonen1998}
\bibinfo{author}{\bibfnamefont{J.}~\bibnamefont{Suhonen}} \bibnamefont{and}
  \bibinfo{author}{\bibfnamefont{O.}~\bibnamefont{Civitarese}},
  \bibinfo{journal}{Physics Reports} \textbf{\bibinfo{volume}{300}},
  \bibinfo{pages}{123} (\bibinfo{year}{1998}), ISSN \bibinfo{issn}{03701573},
  \urlprefix\url{http://dx.doi.org/10.1016/S0370-1573(97)00087-2}.

\bibitem[{\citenamefont{Avignone et~al.}(2008)\citenamefont{Avignone, Elliott,
  and Engel}}]{Avignone2008}
\bibinfo{author}{\bibfnamefont{F.~T.} \bibnamefont{Avignone}},
  \bibinfo{author}{\bibfnamefont{S.~R.} \bibnamefont{Elliott}},
  \bibnamefont{and} \bibinfo{author}{\bibfnamefont{J.}~\bibnamefont{Engel}},
  \bibinfo{journal}{Reviews of Modern Physics} \textbf{\bibinfo{volume}{80}},
  \bibinfo{pages}{481} (\bibinfo{year}{2008}), ISSN \bibinfo{issn}{15390756},
  \eprint{0708.1033},
  \urlprefix\url{http://link.aps.org/doi/10.1103/RevModPhys.80.481}.

\bibitem[{\citenamefont{Ejiri et~al.}(2019)\citenamefont{Ejiri, Suhonen, and
  Zuber}}]{Ejiri2019}
\bibinfo{author}{\bibfnamefont{H.}~\bibnamefont{Ejiri}},
  \bibinfo{author}{\bibfnamefont{J.}~\bibnamefont{Suhonen}}, \bibnamefont{and}
  \bibinfo{author}{\bibfnamefont{K.}~\bibnamefont{Zuber}},
  \bibinfo{journal}{Physics Reports} \textbf{\bibinfo{volume}{797}},
  \bibinfo{pages}{1} (\bibinfo{year}{2019}), ISSN \bibinfo{issn}{03701573},
  \urlprefix\url{https://doi.org/10.1016/j.physrep.2018.12.001
  https://dx.doi.org/10.1016/j.physrep.2018.12.001}.

\bibitem[{\citenamefont{Blaum et~al.}(2020)\citenamefont{Blaum, Eliseev,
  Danevich, Tretyak, Kovalenko, Krivoruchenko, Novikov, and
  Suhonen}}]{Blaum2020}
\bibinfo{author}{\bibfnamefont{K.}~\bibnamefont{Blaum}},
  \bibinfo{author}{\bibfnamefont{S.}~\bibnamefont{Eliseev}},
  \bibinfo{author}{\bibfnamefont{F.~A.} \bibnamefont{Danevich}},
  \bibinfo{author}{\bibfnamefont{V.~I.} \bibnamefont{Tretyak}},
  \bibinfo{author}{\bibfnamefont{S.}~\bibnamefont{Kovalenko}},
  \bibinfo{author}{\bibfnamefont{M.~I.} \bibnamefont{Krivoruchenko}},
  \bibinfo{author}{\bibfnamefont{Y.~N.} \bibnamefont{Novikov}},
  \bibnamefont{and} \bibinfo{author}{\bibfnamefont{J.}~\bibnamefont{Suhonen}},
  \bibinfo{journal}{Reviews of Modern Physics} \textbf{\bibinfo{volume}{92}},
  \bibinfo{pages}{045007} (\bibinfo{year}{2020}),
  \urlprefix\url{https://link.aps.org/doi/10.1103/RevModPhys.92.045007}.

\bibitem[{\citenamefont{Gastaldo et~al.}(2017)\citenamefont{Gastaldo, Blaum,
  Chrysalidis, {Day Goodacre}, Domula, Door, Dorrer, D{\"{u}}llmann, Eberhardt,
  Eliseev et~al.}}]{Gastaldo2017}
\bibinfo{author}{\bibfnamefont{L.}~\bibnamefont{Gastaldo}},
  \bibinfo{author}{\bibfnamefont{K.}~\bibnamefont{Blaum}},
  \bibinfo{author}{\bibfnamefont{K.}~\bibnamefont{Chrysalidis}},
  \bibinfo{author}{\bibfnamefont{T.}~\bibnamefont{{Day Goodacre}}},
  \bibinfo{author}{\bibfnamefont{A.}~\bibnamefont{Domula}},
  \bibinfo{author}{\bibfnamefont{M.}~\bibnamefont{Door}},
  \bibinfo{author}{\bibfnamefont{H.}~\bibnamefont{Dorrer}},
  \bibinfo{author}{\bibfnamefont{C.~E.} \bibnamefont{D{\"{u}}llmann}},
  \bibinfo{author}{\bibfnamefont{K.}~\bibnamefont{Eberhardt}},
  \bibinfo{author}{\bibfnamefont{S.}~\bibnamefont{Eliseev}},
  \bibnamefont{et~al.}, \bibinfo{journal}{European Physical Journal: Special
  Topics} \textbf{\bibinfo{volume}{226}}, \bibinfo{pages}{1623}
  (\bibinfo{year}{2017}), ISSN \bibinfo{issn}{19516401},
  \urlprefix\url{http://link.springer.com/10.1140/epjst/e2017-70071-y}.

\bibitem[{\citenamefont{Aker et~al.}(2021)\citenamefont{Aker, Beglarian,
  Behrens, Berlev, Besserer, Bieringer, Block, Bornschein, Bornschein,
  Böttcher et~al.}}]{aker2021direct}
\bibinfo{author}{\bibfnamefont{M.}~\bibnamefont{Aker}},
  \bibinfo{author}{\bibfnamefont{A.}~\bibnamefont{Beglarian}},
  \bibinfo{author}{\bibfnamefont{J.}~\bibnamefont{Behrens}},
  \bibinfo{author}{\bibfnamefont{A.}~\bibnamefont{Berlev}},
  \bibinfo{author}{\bibfnamefont{U.}~\bibnamefont{Besserer}},
  \bibinfo{author}{\bibfnamefont{B.}~\bibnamefont{Bieringer}},
  \bibinfo{author}{\bibfnamefont{F.}~\bibnamefont{Block}},
  \bibinfo{author}{\bibfnamefont{B.}~\bibnamefont{Bornschein}},
  \bibinfo{author}{\bibfnamefont{L.}~\bibnamefont{Bornschein}},
  \bibinfo{author}{\bibfnamefont{M.}~\bibnamefont{Böttcher}},
  \bibnamefont{et~al.}, \emph{\bibinfo{title}{First direct neutrino-mass
  measurement with sub-{eV} sensitivity}} (\bibinfo{year}{2021}),
  \eprint{2105.08533}.

\bibitem[{\citenamefont{Myers et~al.}(2015)\citenamefont{Myers, Wagner, Kracke,
  and Wesson}}]{Myers2015}
\bibinfo{author}{\bibfnamefont{E.~G.} \bibnamefont{Myers}},
  \bibinfo{author}{\bibfnamefont{A.}~\bibnamefont{Wagner}},
  \bibinfo{author}{\bibfnamefont{H.}~\bibnamefont{Kracke}}, \bibnamefont{and}
  \bibinfo{author}{\bibfnamefont{B.~A.} \bibnamefont{Wesson}},
  \bibinfo{journal}{Physical Review Letters} \textbf{\bibinfo{volume}{114}}
  (\bibinfo{year}{2015}), ISSN \bibinfo{issn}{10797114},
  \urlprefix\url{http://dx.doi.org/10.1103/PhysRevLett.114.013003}.

\bibitem[{\citenamefont{Drexlin et~al.}(2013)\citenamefont{Drexlin, Hannen,
  Mertens, and Weinheimer}}]{Drexlin2013}
\bibinfo{author}{\bibfnamefont{G.}~\bibnamefont{Drexlin}},
  \bibinfo{author}{\bibfnamefont{V.}~\bibnamefont{Hannen}},
  \bibinfo{author}{\bibfnamefont{S.}~\bibnamefont{Mertens}}, \bibnamefont{and}
  \bibinfo{author}{\bibfnamefont{C.}~\bibnamefont{Weinheimer}},
  \bibinfo{journal}{Advances in High Energy Physics}
  \textbf{\bibinfo{volume}{2013}}, \bibinfo{pages}{1} (\bibinfo{year}{2013}),
  ISSN \bibinfo{issn}{16877357}, \eprint{1307.0101},
  \urlprefix\url{http://www.hindawi.com/journals/ahep/2013/293986/}.

\bibitem[{\citenamefont{Aker et~al.}(2019)\citenamefont{Aker,
  Altenm{\"{u}}ller, Arenz, Babutzka, Barrett, Bauer, Beck, Beglarian, Behrens,
  Bergmann et~al.}}]{Aker2019}
\bibinfo{author}{\bibfnamefont{M.}~\bibnamefont{Aker}},
  \bibinfo{author}{\bibfnamefont{K.}~\bibnamefont{Altenm{\"{u}}ller}},
  \bibinfo{author}{\bibfnamefont{M.}~\bibnamefont{Arenz}},
  \bibinfo{author}{\bibfnamefont{M.}~\bibnamefont{Babutzka}},
  \bibinfo{author}{\bibfnamefont{J.}~\bibnamefont{Barrett}},
  \bibinfo{author}{\bibfnamefont{S.}~\bibnamefont{Bauer}},
  \bibinfo{author}{\bibfnamefont{M.}~\bibnamefont{Beck}},
  \bibinfo{author}{\bibfnamefont{A.}~\bibnamefont{Beglarian}},
  \bibinfo{author}{\bibfnamefont{J.}~\bibnamefont{Behrens}},
  \bibinfo{author}{\bibfnamefont{T.}~\bibnamefont{Bergmann}},
  \bibnamefont{et~al.}, \bibinfo{journal}{Physical Review Letters}
  \textbf{\bibinfo{volume}{123}}, \bibinfo{pages}{1} (\bibinfo{year}{2019}),
  ISSN \bibinfo{issn}{10797114}, \eprint{1909.06048},
  \urlprefix\url{https://doi.org/10.1103/PhysRevLett.123.221802}.

\bibitem[{\citenamefont{Eliseev et~al.}(2015)\citenamefont{Eliseev, Blaum,
  Block, Chenmarev, Dorrer, D{\"{u}}llmann, Enss, Filianin, Gastaldo, Goncharov
  et~al.}}]{Eliseev2015}
\bibinfo{author}{\bibfnamefont{S.}~\bibnamefont{Eliseev}},
  \bibinfo{author}{\bibfnamefont{K.}~\bibnamefont{Blaum}},
  \bibinfo{author}{\bibfnamefont{M.}~\bibnamefont{Block}},
  \bibinfo{author}{\bibfnamefont{S.}~\bibnamefont{Chenmarev}},
  \bibinfo{author}{\bibfnamefont{H.}~\bibnamefont{Dorrer}},
  \bibinfo{author}{\bibfnamefont{C.~E.} \bibnamefont{D{\"{u}}llmann}},
  \bibinfo{author}{\bibfnamefont{C.}~\bibnamefont{Enss}},
  \bibinfo{author}{\bibfnamefont{P.~E.} \bibnamefont{Filianin}},
  \bibinfo{author}{\bibfnamefont{L.}~\bibnamefont{Gastaldo}},
  \bibinfo{author}{\bibfnamefont{M.}~\bibnamefont{Goncharov}},
  \bibnamefont{et~al.}, \bibinfo{journal}{Physical Review Letters}
  \textbf{\bibinfo{volume}{115}}, \bibinfo{pages}{62501}
  (\bibinfo{year}{2015}), ISSN \bibinfo{issn}{10797114}, \eprint{1604.04210},
  \urlprefix\url{http://link.aps.org/doi/10.1103/PhysRevLett.115.062501}.

\bibitem[{\citenamefont{Gastaldo et~al.}(2014)\citenamefont{Gastaldo, Blaum,
  Doerr, D{\"{u}}llmann, Eberhardt, Eliseev, Enss, Faessler, Fleischmann, Kempf
  et~al.}}]{Gastaldo2014}
\bibinfo{author}{\bibfnamefont{L.}~\bibnamefont{Gastaldo}},
  \bibinfo{author}{\bibfnamefont{K.}~\bibnamefont{Blaum}},
  \bibinfo{author}{\bibfnamefont{A.}~\bibnamefont{Doerr}},
  \bibinfo{author}{\bibfnamefont{C.~E.} \bibnamefont{D{\"{u}}llmann}},
  \bibinfo{author}{\bibfnamefont{K.}~\bibnamefont{Eberhardt}},
  \bibinfo{author}{\bibfnamefont{S.}~\bibnamefont{Eliseev}},
  \bibinfo{author}{\bibfnamefont{C.}~\bibnamefont{Enss}},
  \bibinfo{author}{\bibfnamefont{A.}~\bibnamefont{Faessler}},
  \bibinfo{author}{\bibfnamefont{A.}~\bibnamefont{Fleischmann}},
  \bibinfo{author}{\bibfnamefont{S.}~\bibnamefont{Kempf}},
  \bibnamefont{et~al.}, \bibinfo{journal}{Journal of Low Temperature Physics}
  \textbf{\bibinfo{volume}{176}}, \bibinfo{pages}{876} (\bibinfo{year}{2014}),
  ISSN \bibinfo{issn}{15737357}, \eprint{1306.2655},
  \urlprefix\url{http://dx.doi.org/10.1007/s10909-014-1187-4}.

\bibitem[{\citenamefont{Alpert et~al.}(2015)\citenamefont{Alpert, Balata,
  Bennett, Biasotti, Boragno, Brofferio, Ceriale, Corsini, Day, {De Gerone}
  et~al.}}]{Alpert2015}
\bibinfo{author}{\bibfnamefont{B.}~\bibnamefont{Alpert}},
  \bibinfo{author}{\bibfnamefont{M.}~\bibnamefont{Balata}},
  \bibinfo{author}{\bibfnamefont{D.}~\bibnamefont{Bennett}},
  \bibinfo{author}{\bibfnamefont{M.}~\bibnamefont{Biasotti}},
  \bibinfo{author}{\bibfnamefont{C.}~\bibnamefont{Boragno}},
  \bibinfo{author}{\bibfnamefont{C.}~\bibnamefont{Brofferio}},
  \bibinfo{author}{\bibfnamefont{V.}~\bibnamefont{Ceriale}},
  \bibinfo{author}{\bibfnamefont{D.}~\bibnamefont{Corsini}},
  \bibinfo{author}{\bibfnamefont{P.~K.} \bibnamefont{Day}},
  \bibinfo{author}{\bibfnamefont{M.}~\bibnamefont{{De Gerone}}},
  \bibnamefont{et~al.}, \bibinfo{journal}{European Physical Journal C}
  \textbf{\bibinfo{volume}{75}}, \bibinfo{pages}{1} (\bibinfo{year}{2015}),
  ISSN \bibinfo{issn}{14346052}, \eprint{1412.5060},
  \urlprefix\url{http://dx.doi.org/10.1140/epjc/s10052-015-3329-5
  https://www.ncbi.nlm.nih.gov/pubmed/25995704}.

\bibitem[{\citenamefont{Faverzani et~al.}(2016)\citenamefont{Faverzani, Alpert,
  Backer, Bennet, Biasotti, Brofferio, Ceriale, Ceruti, Corsini, Day
  et~al.}}]{Faverzani2016}
\bibinfo{author}{\bibfnamefont{M.}~\bibnamefont{Faverzani}},
  \bibinfo{author}{\bibfnamefont{B.}~\bibnamefont{Alpert}},
  \bibinfo{author}{\bibfnamefont{D.}~\bibnamefont{Backer}},
  \bibinfo{author}{\bibfnamefont{D.}~\bibnamefont{Bennet}},
  \bibinfo{author}{\bibfnamefont{M.}~\bibnamefont{Biasotti}},
  \bibinfo{author}{\bibfnamefont{C.}~\bibnamefont{Brofferio}},
  \bibinfo{author}{\bibfnamefont{V.}~\bibnamefont{Ceriale}},
  \bibinfo{author}{\bibfnamefont{G.}~\bibnamefont{Ceruti}},
  \bibinfo{author}{\bibfnamefont{D.}~\bibnamefont{Corsini}},
  \bibinfo{author}{\bibfnamefont{P.~K.} \bibnamefont{Day}},
  \bibnamefont{et~al.}, \bibinfo{journal}{Journal of Low Temperature Physics}
  \textbf{\bibinfo{volume}{184}}, \bibinfo{pages}{922} (\bibinfo{year}{2016}),
  ISSN \bibinfo{issn}{15737357},
  \urlprefix\url{http://link.springer.com/10.1007/s10909-016-1540-x}.

\bibitem[{\citenamefont{Velte et~al.}(2019)\citenamefont{Velte, Ahrens, Barth,
  Blaum, Bra{\ss}, Door, Dorrer, Düllmann, Eliseev, Enss et~al.}}]{Velte2019}
\bibinfo{author}{\bibfnamefont{C.}~\bibnamefont{Velte}},
  \bibinfo{author}{\bibfnamefont{F.}~\bibnamefont{Ahrens}},
  \bibinfo{author}{\bibfnamefont{A.}~\bibnamefont{Barth}},
  \bibinfo{author}{\bibfnamefont{K.}~\bibnamefont{Blaum}},
  \bibinfo{author}{\bibfnamefont{M.}~\bibnamefont{Bra{\ss}}},
  \bibinfo{author}{\bibfnamefont{M.}~\bibnamefont{Door}},
  \bibinfo{author}{\bibfnamefont{H.}~\bibnamefont{Dorrer}},
  \bibinfo{author}{\bibfnamefont{C.~E.} \bibnamefont{Düllmann}},
  \bibinfo{author}{\bibfnamefont{S.}~\bibnamefont{Eliseev}},
  \bibinfo{author}{\bibfnamefont{C.}~\bibnamefont{Enss}}, \bibnamefont{et~al.},
  \bibinfo{journal}{The European Physical Journal C}
  \textbf{\bibinfo{volume}{79}} (\bibinfo{year}{2019}),
  \urlprefix\url{http://link.springer.com/10.1140/epjc/s10052-019-7513-x}.

\bibitem[{\citenamefont{Mustonen and Suhonen}(2010)}]{Mustonen2010}
\bibinfo{author}{\bibfnamefont{M.~T.} \bibnamefont{Mustonen}} \bibnamefont{and}
  \bibinfo{author}{\bibfnamefont{J.}~\bibnamefont{Suhonen}},
  \bibinfo{journal}{Journal of Physics G: Nuclear and Particle Physics}
  \textbf{\bibinfo{volume}{37}}, \bibinfo{pages}{64008} (\bibinfo{year}{2010}),
  ISSN \bibinfo{issn}{09543899},
  \urlprefix\url{https://iopscience.iop.org/article/10.1088/0954-3899/37/6/064008}.

\bibitem[{\citenamefont{Mustonen and Suhonen}(2011)}]{Mustonen2011}
\bibinfo{author}{\bibfnamefont{M.}~\bibnamefont{Mustonen}} \bibnamefont{and}
  \bibinfo{author}{\bibfnamefont{J.}~\bibnamefont{Suhonen}},
  \bibinfo{journal}{Physics Letters B} \textbf{\bibinfo{volume}{703}},
  \bibinfo{pages}{370} (\bibinfo{year}{2011}), ISSN \bibinfo{issn}{0370-2693},
  \urlprefix\url{https://www.sciencedirect.com/science/article/pii/S0370269311009099}.

\bibitem[{\citenamefont{Suhonen}(2014)}]{Suhonen2014}
\bibinfo{author}{\bibfnamefont{J.}~\bibnamefont{Suhonen}},
  \bibinfo{journal}{Physica Scripta} \textbf{\bibinfo{volume}{89}},
  \bibinfo{pages}{54032} (\bibinfo{year}{2014}), ISSN \bibinfo{issn}{14024896},
  \urlprefix\url{http://stacks.iop.org/1402-4896/89/i=5/a=054032}.

\bibitem[{\citenamefont{{De Roubin} et~al.}(2020)\citenamefont{{De Roubin},
  Kostensalo, Eronen, Canete, {De Groote}, Jokinen, Kankainen, Nesterenko,
  Moore, Rinta-Antila et~al.}}]{deRoubin2020}
\bibinfo{author}{\bibfnamefont{A.}~\bibnamefont{{De Roubin}}},
  \bibinfo{author}{\bibfnamefont{J.}~\bibnamefont{Kostensalo}},
  \bibinfo{author}{\bibfnamefont{T.}~\bibnamefont{Eronen}},
  \bibinfo{author}{\bibfnamefont{L.}~\bibnamefont{Canete}},
  \bibinfo{author}{\bibfnamefont{R.~P.} \bibnamefont{{De Groote}}},
  \bibinfo{author}{\bibfnamefont{A.}~\bibnamefont{Jokinen}},
  \bibinfo{author}{\bibfnamefont{A.}~\bibnamefont{Kankainen}},
  \bibinfo{author}{\bibfnamefont{D.~A.} \bibnamefont{Nesterenko}},
  \bibinfo{author}{\bibfnamefont{I.~D.} \bibnamefont{Moore}},
  \bibinfo{author}{\bibfnamefont{S.}~\bibnamefont{Rinta-Antila}},
  \bibnamefont{et~al.}, \bibinfo{journal}{Physical Review Letters}
  \textbf{\bibinfo{volume}{124}}, \bibinfo{pages}{1} (\bibinfo{year}{2020}),
  ISSN \bibinfo{issn}{10797114}, \eprint{2002.08282},
  \urlprefix\url{https://doi.org/10.1103/PhysRevLett.124.222503}.

\bibitem[{\citenamefont{Ge et~al.}(2021{\natexlab{a}})\citenamefont{Ge, Eronen,
  de~Roubin, Nesterenko, Hukkanen, Beliuskina, de~Groote, Geldhof, Gins,
  Kankainen et~al.}}]{ge2021}
\bibinfo{author}{\bibfnamefont{Z.}~\bibnamefont{Ge}},
  \bibinfo{author}{\bibfnamefont{T.}~\bibnamefont{Eronen}},
  \bibinfo{author}{\bibfnamefont{A.}~\bibnamefont{de~Roubin}},
  \bibinfo{author}{\bibfnamefont{D.~A.} \bibnamefont{Nesterenko}},
  \bibinfo{author}{\bibfnamefont{M.}~\bibnamefont{Hukkanen}},
  \bibinfo{author}{\bibfnamefont{O.}~\bibnamefont{Beliuskina}},
  \bibinfo{author}{\bibfnamefont{R.}~\bibnamefont{de~Groote}},
  \bibinfo{author}{\bibfnamefont{S.}~\bibnamefont{Geldhof}},
  \bibinfo{author}{\bibfnamefont{W.}~\bibnamefont{Gins}},
  \bibinfo{author}{\bibfnamefont{A.}~\bibnamefont{Kankainen}},
  \bibnamefont{et~al.}, \bibinfo{journal}{Physical Review C}
  \textbf{\bibinfo{volume}{103}}, \bibinfo{pages}{065502}
  (\bibinfo{year}{2021}{\natexlab{a}}),
  \urlprefix\url{https://link.aps.org/doi/10.1103/PhysRevC.103.065502}.

\bibitem[{\citenamefont{Ge et~al.}(2021{\natexlab{b}})\citenamefont{Ge, Eronen,
  Tyrin, Kotila, Kostensalo, Nesterenko, Beliuskina, de~Groote, de~Roubin,
  Geldhof et~al.}}]{ge2021b}
\bibinfo{author}{\bibfnamefont{Z.}~\bibnamefont{Ge}},
  \bibinfo{author}{\bibfnamefont{T.}~\bibnamefont{Eronen}},
  \bibinfo{author}{\bibfnamefont{K.~S.} \bibnamefont{Tyrin}},
  \bibinfo{author}{\bibfnamefont{J.}~\bibnamefont{Kotila}},
  \bibinfo{author}{\bibfnamefont{J.}~\bibnamefont{Kostensalo}},
  \bibinfo{author}{\bibfnamefont{D.~A.} \bibnamefont{Nesterenko}},
  \bibinfo{author}{\bibfnamefont{O.}~\bibnamefont{Beliuskina}},
  \bibinfo{author}{\bibfnamefont{R.}~\bibnamefont{de~Groote}},
  \bibinfo{author}{\bibfnamefont{A.}~\bibnamefont{de~Roubin}},
  \bibinfo{author}{\bibfnamefont{S.}~\bibnamefont{Geldhof}},
  \bibnamefont{et~al.}, \bibinfo{journal}{Phys. Rev. Lett.}
  \textbf{\bibinfo{volume}{127}}, \bibinfo{pages}{272301}
  (\bibinfo{year}{2021}{\natexlab{b}}),
  \urlprefix\url{https://link.aps.org/doi/10.1103/PhysRevLett.127.272301}.

\bibitem[{\citenamefont{Fink et~al.}(2012)\citenamefont{Fink, Barea, Beck,
  Blaum, B{\"{o}}hm, Borgmann, Breitenfeldt, Herfurth, Herlert, Kotila
  et~al.}}]{Fink2012}
\bibinfo{author}{\bibfnamefont{D.}~\bibnamefont{Fink}},
  \bibinfo{author}{\bibfnamefont{J.}~\bibnamefont{Barea}},
  \bibinfo{author}{\bibfnamefont{D.}~\bibnamefont{Beck}},
  \bibinfo{author}{\bibfnamefont{K.}~\bibnamefont{Blaum}},
  \bibinfo{author}{\bibfnamefont{C.}~\bibnamefont{B{\"{o}}hm}},
  \bibinfo{author}{\bibfnamefont{C.}~\bibnamefont{Borgmann}},
  \bibinfo{author}{\bibfnamefont{M.}~\bibnamefont{Breitenfeldt}},
  \bibinfo{author}{\bibfnamefont{F.}~\bibnamefont{Herfurth}},
  \bibinfo{author}{\bibfnamefont{A.}~\bibnamefont{Herlert}},
  \bibinfo{author}{\bibfnamefont{J.}~\bibnamefont{Kotila}},
  \bibnamefont{et~al.}, \bibinfo{journal}{Physical Review Letters}
  \textbf{\bibinfo{volume}{108}}, \bibinfo{pages}{1} (\bibinfo{year}{2012}),
  ISSN \bibinfo{issn}{00319007}.

\bibitem[{\citenamefont{Nesterenko et~al.}(2019)\citenamefont{Nesterenko,
  Canete, Eronen, Jokinen, Kankainen, Novikov, Rinta-Antila, de~Roubin, and
  Vilen}}]{Nesterenko2019}
\bibinfo{author}{\bibfnamefont{D.~A.} \bibnamefont{Nesterenko}},
  \bibinfo{author}{\bibfnamefont{L.}~\bibnamefont{Canete}},
  \bibinfo{author}{\bibfnamefont{T.}~\bibnamefont{Eronen}},
  \bibinfo{author}{\bibfnamefont{A.}~\bibnamefont{Jokinen}},
  \bibinfo{author}{\bibfnamefont{A.}~\bibnamefont{Kankainen}},
  \bibinfo{author}{\bibfnamefont{Y.~N.} \bibnamefont{Novikov}},
  \bibinfo{author}{\bibfnamefont{S.}~\bibnamefont{Rinta-Antila}},
  \bibinfo{author}{\bibfnamefont{A.}~\bibnamefont{de~Roubin}},
  \bibnamefont{and} \bibinfo{author}{\bibfnamefont{M.}~\bibnamefont{Vilen}},
  \bibinfo{journal}{International Journal of Mass Spectrometry}
  \textbf{\bibinfo{volume}{435}}, \bibinfo{pages}{204} (\bibinfo{year}{2019}),
  ISSN \bibinfo{issn}{13873806},
  \urlprefix\url{https://dx.doi.org/10.1016/j.ijms.2018.10.038
  https://doi.org/10.1016/j.ijms.2018.10.038}.

\bibitem[{\citenamefont{Cattadori et~al.}(2005)\citenamefont{Cattadori, {De
  Deo}, Laubenstein, Pandola, and Tretyak}}]{Cattadori2005}
\bibinfo{author}{\bibfnamefont{C.~M.} \bibnamefont{Cattadori}},
  \bibinfo{author}{\bibfnamefont{M.}~\bibnamefont{{De Deo}}},
  \bibinfo{author}{\bibfnamefont{M.}~\bibnamefont{Laubenstein}},
  \bibinfo{author}{\bibfnamefont{L.}~\bibnamefont{Pandola}}, \bibnamefont{and}
  \bibinfo{author}{\bibfnamefont{V.~I.} \bibnamefont{Tretyak}},
  \bibinfo{journal}{Nuclear Physics A} \textbf{\bibinfo{volume}{748}},
  \bibinfo{pages}{333} (\bibinfo{year}{2005}), ISSN \bibinfo{issn}{03759474},
  \urlprefix\url{http://www.sciencedirect.com/science/article/pii/S0375947404011509}.

\bibitem[{\citenamefont{Wieslander et~al.}(2009)\citenamefont{Wieslander,
  Suhonen, Eronen, Hult, Elomaa, Jokinen, Marissens, Misiaszek, Mustonen,
  Rahaman et~al.}}]{Wieslander2009}
\bibinfo{author}{\bibfnamefont{J.~S.} \bibnamefont{Wieslander}},
  \bibinfo{author}{\bibfnamefont{J.}~\bibnamefont{Suhonen}},
  \bibinfo{author}{\bibfnamefont{T.}~\bibnamefont{Eronen}},
  \bibinfo{author}{\bibfnamefont{M.}~\bibnamefont{Hult}},
  \bibinfo{author}{\bibfnamefont{V.~V.} \bibnamefont{Elomaa}},
  \bibinfo{author}{\bibfnamefont{A.}~\bibnamefont{Jokinen}},
  \bibinfo{author}{\bibfnamefont{G.}~\bibnamefont{Marissens}},
  \bibinfo{author}{\bibfnamefont{M.}~\bibnamefont{Misiaszek}},
  \bibinfo{author}{\bibfnamefont{M.~T.} \bibnamefont{Mustonen}},
  \bibinfo{author}{\bibfnamefont{S.}~\bibnamefont{Rahaman}},
  \bibnamefont{et~al.}, \bibinfo{journal}{Physical Review Letters}
  \textbf{\bibinfo{volume}{103}}, \bibinfo{pages}{122501}
  (\bibinfo{year}{2009}), ISSN \bibinfo{issn}{00319007},
  \urlprefix\url{https://www.ncbi.nlm.nih.gov/pubmed/19792426
  http://dx.doi.org/10.1103/PhysRevLett.103.122501
  http://link.aps.org/doi/10.1103/PhysRevLett.103.122501}.

\bibitem[{\citenamefont{Mount et~al.}(2009)\citenamefont{Mount, Redshaw, and
  Myers}}]{Mount2009}
\bibinfo{author}{\bibfnamefont{B.~J.} \bibnamefont{Mount}},
  \bibinfo{author}{\bibfnamefont{M.}~\bibnamefont{Redshaw}}, \bibnamefont{and}
  \bibinfo{author}{\bibfnamefont{E.~G.} \bibnamefont{Myers}},
  \bibinfo{journal}{Physical Review Letters} \textbf{\bibinfo{volume}{103}},
  \bibinfo{pages}{122502} (\bibinfo{year}{2009}), ISSN
  \bibinfo{issn}{00319007},
  \urlprefix\url{http://dx.doi.org/10.1103/PhysRevLett.103.122502
  http://link.aps.org/doi/10.1103/PhysRevLett.103.122502}.

\bibitem[{\citenamefont{Kankainen et~al.}(2020)\citenamefont{Kankainen, Eronen,
  Nesterenko, de~Roubin, and Vil{\'{e}}n}}]{Kankainen2020}
\bibinfo{author}{\bibfnamefont{A.}~\bibnamefont{Kankainen}},
  \bibinfo{author}{\bibfnamefont{T.}~\bibnamefont{Eronen}},
  \bibinfo{author}{\bibfnamefont{D.}~\bibnamefont{Nesterenko}},
  \bibinfo{author}{\bibfnamefont{A.}~\bibnamefont{de~Roubin}},
  \bibnamefont{and}
  \bibinfo{author}{\bibfnamefont{M.}~\bibnamefont{Vil{\'{e}}n}},
  \bibinfo{journal}{Hyperfine Interactions} \textbf{\bibinfo{volume}{241}},
  \bibinfo{pages}{43} (\bibinfo{year}{2020}), ISSN \bibinfo{issn}{15729540},
  \urlprefix\url{http://link.springer.com/10.1007/s10751-020-01711-5}.

\bibitem[{\citenamefont{Eronen and Hardy}(2012)}]{Eronen2012}
\bibinfo{author}{\bibfnamefont{T.}~\bibnamefont{Eronen}} \bibnamefont{and}
  \bibinfo{author}{\bibfnamefont{J.~C.} \bibnamefont{Hardy}},
  \bibinfo{journal}{European Physical Journal A} \textbf{\bibinfo{volume}{48}},
  \bibinfo{pages}{1} (\bibinfo{year}{2012}), ISSN \bibinfo{issn}{1434601X},
  \urlprefix\url{http://dx.doi.org/10.1140/epja/i2012-12048-y}.

\bibitem[{\citenamefont{Moore et~al.}(2013)\citenamefont{Moore, Eronen,
  Gorelov, Hakala, Jokinen, Kankainen, Kolhinen, Koponen, Penttil{\"{a}},
  Pohjalainen et~al.}}]{Moore2013}
\bibinfo{author}{\bibfnamefont{I.~D.} \bibnamefont{Moore}},
  \bibinfo{author}{\bibfnamefont{T.}~\bibnamefont{Eronen}},
  \bibinfo{author}{\bibfnamefont{D.}~\bibnamefont{Gorelov}},
  \bibinfo{author}{\bibfnamefont{J.}~\bibnamefont{Hakala}},
  \bibinfo{author}{\bibfnamefont{A.}~\bibnamefont{Jokinen}},
  \bibinfo{author}{\bibfnamefont{A.}~\bibnamefont{Kankainen}},
  \bibinfo{author}{\bibfnamefont{V.~S.} \bibnamefont{Kolhinen}},
  \bibinfo{author}{\bibfnamefont{J.}~\bibnamefont{Koponen}},
  \bibinfo{author}{\bibfnamefont{H.}~\bibnamefont{Penttil{\"{a}}}},
  \bibinfo{author}{\bibfnamefont{I.}~\bibnamefont{Pohjalainen}},
  \bibnamefont{et~al.}, \bibinfo{journal}{Nuclear Instruments and Methods in
  Physics Research, Section B: Beam Interactions with Materials and Atoms}
  \textbf{\bibinfo{volume}{317}}, \bibinfo{pages}{208} (\bibinfo{year}{2013}),
  ISSN \bibinfo{issn}{0168583X},
  \urlprefix\url{http://www.sciencedirect.com/science/article/pii/S0168583X13007143
  http://dx.doi.org/10.1016/j.nimb.2013.06.036}.

\bibitem[{\citenamefont{Nieminen et~al.}(2001)\citenamefont{Nieminen, Huikari,
  Jokinen, {\"{A}}yst{\"{o}}, Campbell, and Cochrane}}]{Nieminen2001}
\bibinfo{author}{\bibfnamefont{A.}~\bibnamefont{Nieminen}},
  \bibinfo{author}{\bibfnamefont{J.}~\bibnamefont{Huikari}},
  \bibinfo{author}{\bibfnamefont{A.}~\bibnamefont{Jokinen}},
  \bibinfo{author}{\bibfnamefont{J.}~\bibnamefont{{\"{A}}yst{\"{o}}}},
  \bibinfo{author}{\bibfnamefont{P.}~\bibnamefont{Campbell}}, \bibnamefont{and}
  \bibinfo{author}{\bibfnamefont{E.~C.} \bibnamefont{Cochrane}},
  \bibinfo{journal}{Nuclear Instruments and Methods in Physics Research,
  Section A: Accelerators, Spectrometers, Detectors and Associated Equipment}
  \textbf{\bibinfo{volume}{469}}, \bibinfo{pages}{244} (\bibinfo{year}{2001}),
  ISSN \bibinfo{issn}{01689002},
  \urlprefix\url{http://www.sciencedirect.com/science/article/B6TJM-43PGJKX-C/1/93d5587efba5cfe8571b63228952dab8}.

\bibitem[{\citenamefont{Savard et~al.}(1991)\citenamefont{Savard, Becker,
  Bollen, Kluge, Moore, Otto, Schweikhard, Stolzenberg, and
  Wiess}}]{Savard1991}
\bibinfo{author}{\bibfnamefont{G.}~\bibnamefont{Savard}},
  \bibinfo{author}{\bibfnamefont{S.}~\bibnamefont{Becker}},
  \bibinfo{author}{\bibfnamefont{G.}~\bibnamefont{Bollen}},
  \bibinfo{author}{\bibfnamefont{H.~J.} \bibnamefont{Kluge}},
  \bibinfo{author}{\bibfnamefont{R.~B.} \bibnamefont{Moore}},
  \bibinfo{author}{\bibfnamefont{T.}~\bibnamefont{Otto}},
  \bibinfo{author}{\bibfnamefont{L.}~\bibnamefont{Schweikhard}},
  \bibinfo{author}{\bibfnamefont{H.}~\bibnamefont{Stolzenberg}},
  \bibnamefont{and} \bibinfo{author}{\bibfnamefont{U.}~\bibnamefont{Wiess}},
  \bibinfo{journal}{Physics Letters A} \textbf{\bibinfo{volume}{158}},
  \bibinfo{pages}{247} (\bibinfo{year}{1991}), ISSN \bibinfo{issn}{03759601},
  \urlprefix\url{https://doi.org/10.1016/0375-9601(91)91008-2}.

\bibitem[{\citenamefont{Nesterenko et~al.}(2018)\citenamefont{Nesterenko,
  Eronen, Kankainen, Canete, Jokinen, Moore, Penttil{\"{a}}, Rinta-Antila,
  de~Roubin, and Vilen}}]{Nesterenko2018}
\bibinfo{author}{\bibfnamefont{D.~A.} \bibnamefont{Nesterenko}},
  \bibinfo{author}{\bibfnamefont{T.}~\bibnamefont{Eronen}},
  \bibinfo{author}{\bibfnamefont{A.}~\bibnamefont{Kankainen}},
  \bibinfo{author}{\bibfnamefont{L.}~\bibnamefont{Canete}},
  \bibinfo{author}{\bibfnamefont{A.}~\bibnamefont{Jokinen}},
  \bibinfo{author}{\bibfnamefont{I.~D.} \bibnamefont{Moore}},
  \bibinfo{author}{\bibfnamefont{H.}~\bibnamefont{Penttil{\"{a}}}},
  \bibinfo{author}{\bibfnamefont{S.}~\bibnamefont{Rinta-Antila}},
  \bibinfo{author}{\bibfnamefont{A.}~\bibnamefont{de~Roubin}},
  \bibnamefont{and} \bibinfo{author}{\bibfnamefont{M.}~\bibnamefont{Vilen}},
  \bibinfo{journal}{European Physical Journal A} \textbf{\bibinfo{volume}{54}},
  \bibinfo{pages}{0} (\bibinfo{year}{2018}), ISSN \bibinfo{issn}{1434601X},
  \urlprefix\url{https://dx.doi.org/10.1140/epja/i2018-12589-y}.

\bibitem[{\citenamefont{Eliseev et~al.}(2014)\citenamefont{Eliseev, Blaum,
  Block, D{\"{o}}rr, Droese, Eronen, Goncharov, H{\"{o}}cker, Ketter, Ramirez
  et~al.}}]{Eliseev2014}
\bibinfo{author}{\bibfnamefont{S.}~\bibnamefont{Eliseev}},
  \bibinfo{author}{\bibfnamefont{K.}~\bibnamefont{Blaum}},
  \bibinfo{author}{\bibfnamefont{M.}~\bibnamefont{Block}},
  \bibinfo{author}{\bibfnamefont{A.}~\bibnamefont{D{\"{o}}rr}},
  \bibinfo{author}{\bibfnamefont{C.}~\bibnamefont{Droese}},
  \bibinfo{author}{\bibfnamefont{T.}~\bibnamefont{Eronen}},
  \bibinfo{author}{\bibfnamefont{M.}~\bibnamefont{Goncharov}},
  \bibinfo{author}{\bibfnamefont{M.}~\bibnamefont{H{\"{o}}cker}},
  \bibinfo{author}{\bibfnamefont{J.}~\bibnamefont{Ketter}},
  \bibinfo{author}{\bibfnamefont{E.~M.} \bibnamefont{Ramirez}},
  \bibnamefont{et~al.}, \bibinfo{journal}{Applied Physics B: Lasers and Optics}
  \textbf{\bibinfo{volume}{114}}, \bibinfo{pages}{107} (\bibinfo{year}{2014}),
  ISSN \bibinfo{issn}{09462171},
  \urlprefix\url{http://dx.doi.org/10.1007/s00340-013-5621-0}.

\bibitem[{\citenamefont{Eliseev et~al.}(2013)\citenamefont{Eliseev, Blaum,
  Block, Droese, Goncharov, {Minaya Ramirez}, Nesterenko, Novikov, and
  Schweikhard}}]{Eliseev2013}
\bibinfo{author}{\bibfnamefont{S.}~\bibnamefont{Eliseev}},
  \bibinfo{author}{\bibfnamefont{K.}~\bibnamefont{Blaum}},
  \bibinfo{author}{\bibfnamefont{M.}~\bibnamefont{Block}},
  \bibinfo{author}{\bibfnamefont{C.}~\bibnamefont{Droese}},
  \bibinfo{author}{\bibfnamefont{M.}~\bibnamefont{Goncharov}},
  \bibinfo{author}{\bibfnamefont{E.}~\bibnamefont{{Minaya Ramirez}}},
  \bibinfo{author}{\bibfnamefont{D.~A.} \bibnamefont{Nesterenko}},
  \bibinfo{author}{\bibfnamefont{Y.~N.} \bibnamefont{Novikov}},
  \bibnamefont{and}
  \bibinfo{author}{\bibfnamefont{L.}~\bibnamefont{Schweikhard}},
  \bibinfo{journal}{Physical Review Letters} \textbf{\bibinfo{volume}{110}},
  \bibinfo{pages}{82501} (\bibinfo{year}{2013}), ISSN \bibinfo{issn}{00319007},
  \urlprefix\url{http://link.aps.org/doi/10.1103/PhysRevLett.110.082501}.

\bibitem[{PS-(2021)}]{PS-MCP}
\emph{\bibinfo{title}{Micro-channel plate detector with delay line anode,
  roentdek handels gmbh}}, \bibinfo{howpublished}{Available at
  \url{http://www.roentdek.de} (2020/11/30)} (\bibinfo{year}{2021}),
  \urlprefix\url{http://www.roentdek.de}.

\bibitem[{\citenamefont{Kramida et~al.}(2020)\citenamefont{Kramida,
  {Yu.~Ralchenko}, Reader, and {and NIST ASD Team}}}]{NIST_ASD}
\bibinfo{author}{\bibfnamefont{A.}~\bibnamefont{Kramida}},
  \bibinfo{author}{\bibnamefont{{Yu.~Ralchenko}}},
  \bibinfo{author}{\bibfnamefont{J.}~\bibnamefont{Reader}}, \bibnamefont{and}
  \bibinfo{author}{\bibnamefont{{and NIST ASD Team}}},
  \bibinfo{howpublished}{{NIST Atomic Spectra Database (ver. 5.8), [Online].
  Available: {\tt{https://physics.nist.gov/asd}} [2021, January 19]. National
  Institute of Standards and Technology, Gaithersburg, MD.}}
  (\bibinfo{year}{2020}).

\bibitem[{\citenamefont{Wang et~al.}(2021)\citenamefont{Wang, Huang, Kondev,
  Audi, and Naimi}}]{Wang2021}
\bibinfo{author}{\bibfnamefont{M.}~\bibnamefont{Wang}},
  \bibinfo{author}{\bibfnamefont{W.}~\bibnamefont{Huang}},
  \bibinfo{author}{\bibfnamefont{F.}~\bibnamefont{Kondev}},
  \bibinfo{author}{\bibfnamefont{G.}~\bibnamefont{Audi}}, \bibnamefont{and}
  \bibinfo{author}{\bibfnamefont{S.}~\bibnamefont{Naimi}},
  \bibinfo{journal}{Chinese Physics C} \textbf{\bibinfo{volume}{45}},
  \bibinfo{pages}{030003} (\bibinfo{year}{2021}),
  \urlprefix\url{https://doi.org/10.1088/1674-1137/abddaf}.

\bibitem[{NND(2021)}]{NNDC}
\emph{\bibinfo{title}{National nuclear data center}},
  \bibinfo{howpublished}{Available at \url{https://www.nndc.bnl.gov/}
  (2020/4/7)} (\bibinfo{year}{2021}),
  \urlprefix\url{https://www.nndc.bnl.gov/}.

\bibitem[{\citenamefont{Kellerbauer et~al.}(2003)\citenamefont{Kellerbauer,
  Blaum, Bollen, Herfurth, Kluge, Kuckein, Sauvan, Scheidenberger, and
  Schweikhard}}]{Kellerbauer2003}
\bibinfo{author}{\bibfnamefont{A.}~\bibnamefont{Kellerbauer}},
  \bibinfo{author}{\bibfnamefont{K.}~\bibnamefont{Blaum}},
  \bibinfo{author}{\bibfnamefont{G.}~\bibnamefont{Bollen}},
  \bibinfo{author}{\bibfnamefont{F.}~\bibnamefont{Herfurth}},
  \bibinfo{author}{\bibfnamefont{H.~J.} \bibnamefont{Kluge}},
  \bibinfo{author}{\bibfnamefont{M.}~\bibnamefont{Kuckein}},
  \bibinfo{author}{\bibfnamefont{E.}~\bibnamefont{Sauvan}},
  \bibinfo{author}{\bibfnamefont{C.}~\bibnamefont{Scheidenberger}},
  \bibnamefont{and}
  \bibinfo{author}{\bibfnamefont{L.}~\bibnamefont{Schweikhard}},
  \bibinfo{journal}{European Physical Journal D} \textbf{\bibinfo{volume}{22}},
  \bibinfo{pages}{53} (\bibinfo{year}{2003}), ISSN \bibinfo{issn}{14346060},
  \urlprefix\url{http://dx.doi.org/10.1140/epjd/e2002-00222-0}.

\bibitem[{\citenamefont{Roux et~al.}(2013)\citenamefont{Roux, Blaum, Block,
  Droese, Eliseev, Goncharov, Herfurth, Ramirez, Nesterenko, Novikov
  et~al.}}]{Roux2013}
\bibinfo{author}{\bibfnamefont{C.}~\bibnamefont{Roux}},
  \bibinfo{author}{\bibfnamefont{K.}~\bibnamefont{Blaum}},
  \bibinfo{author}{\bibfnamefont{M.}~\bibnamefont{Block}},
  \bibinfo{author}{\bibfnamefont{C.}~\bibnamefont{Droese}},
  \bibinfo{author}{\bibfnamefont{S.}~\bibnamefont{Eliseev}},
  \bibinfo{author}{\bibfnamefont{M.}~\bibnamefont{Goncharov}},
  \bibinfo{author}{\bibfnamefont{F.}~\bibnamefont{Herfurth}},
  \bibinfo{author}{\bibfnamefont{E.~M.} \bibnamefont{Ramirez}},
  \bibinfo{author}{\bibfnamefont{D.~A.} \bibnamefont{Nesterenko}},
  \bibinfo{author}{\bibfnamefont{Y.~N.} \bibnamefont{Novikov}},
  \bibnamefont{et~al.}, \bibinfo{journal}{The European Physical Journal D}
  \textbf{\bibinfo{volume}{67}}, \bibinfo{pages}{1} (\bibinfo{year}{2013}),
  ISSN \bibinfo{issn}{1434-6060},
  \urlprefix\url{http://dx.doi.org/10.1140/epjd/e2013-40110-x}.

\bibitem[{\citenamefont{Nesterenko et~al.}(2021)\citenamefont{Nesterenko,
  Eronen, Ge, Kankainen, and Vilen}}]{nesterenko2021}
\bibinfo{author}{\bibfnamefont{D.~A.} \bibnamefont{Nesterenko}},
  \bibinfo{author}{\bibfnamefont{T.}~\bibnamefont{Eronen}},
  \bibinfo{author}{\bibfnamefont{Z.}~\bibnamefont{Ge}},
  \bibinfo{author}{\bibfnamefont{A.}~\bibnamefont{Kankainen}},
  \bibnamefont{and} \bibinfo{author}{\bibfnamefont{M.}~\bibnamefont{Vilen}},
  \bibinfo{journal}{Eur. Phys. J. A} \textbf{\bibinfo{volume}{57}},
  \bibinfo{pages}{302} (\bibinfo{year}{2021}),
  \urlprefix\url{https://doi.org/10.1140/epja/s10050-021-00608-3}.

\bibitem[{\citenamefont{Nesterenko et~al.}(2020)\citenamefont{Nesterenko,
  de~Groote, Eronen, Ge, Hukkanen, Jokinen, and Kankainen}}]{Nesterenko2020}
\bibinfo{author}{\bibfnamefont{D.~A.} \bibnamefont{Nesterenko}},
  \bibinfo{author}{\bibfnamefont{R.~P.} \bibnamefont{de~Groote}},
  \bibinfo{author}{\bibfnamefont{T.}~\bibnamefont{Eronen}},
  \bibinfo{author}{\bibfnamefont{Z.}~\bibnamefont{Ge}},
  \bibinfo{author}{\bibfnamefont{M.}~\bibnamefont{Hukkanen}},
  \bibinfo{author}{\bibfnamefont{A.}~\bibnamefont{Jokinen}}, \bibnamefont{and}
  \bibinfo{author}{\bibfnamefont{A.}~\bibnamefont{Kankainen}},
  \bibinfo{journal}{International Journal of Mass Spectrometry}
  \textbf{\bibinfo{volume}{458}}, \bibinfo{pages}{1} (\bibinfo{year}{2020}),
  ISSN \bibinfo{issn}{13873806}, \eprint{2007.13375}.

\bibitem[{\citenamefont{Huang et~al.}(2021)\citenamefont{Huang, Wang, Kondev,
  Audi, and Naimi}}]{Huang2021}
\bibinfo{author}{\bibfnamefont{W.}~\bibnamefont{Huang}},
  \bibinfo{author}{\bibfnamefont{M.}~\bibnamefont{Wang}},
  \bibinfo{author}{\bibfnamefont{F.}~\bibnamefont{Kondev}},
  \bibinfo{author}{\bibfnamefont{G.}~\bibnamefont{Audi}}, \bibnamefont{and}
  \bibinfo{author}{\bibfnamefont{S.}~\bibnamefont{Naimi}},
  \bibinfo{journal}{Chinese Physics C} \textbf{\bibinfo{volume}{45}},
  \bibinfo{pages}{030002} (\bibinfo{year}{2021}),
  \urlprefix\url{https://doi.org/10.1088/1674-1137/abddb0}.

\bibitem[{\citenamefont{Birge}(1932)}]{Birge1932}
\bibinfo{author}{\bibfnamefont{R.~T.} \bibnamefont{Birge}},
  \bibinfo{journal}{Physical Review} \textbf{\bibinfo{volume}{40}},
  \bibinfo{pages}{207} (\bibinfo{year}{1932}), ISSN \bibinfo{issn}{0031899X},
  \urlprefix\url{http://link.aps.org/abstract/PR/v40/p207}.

\bibitem[{\citenamefont{Behrens and B{\"{u}}hring}(1982)}]{Behrens1982}
\bibinfo{author}{\bibfnamefont{H.}~\bibnamefont{Behrens}} \bibnamefont{and}
  \bibinfo{author}{\bibfnamefont{W.}~\bibnamefont{B{\"{u}}hring}},
  \emph{\bibinfo{title}{{Electron Radial Wave Functions and Nuclear Beta-decay
  (International Series of Monographs on Physics)}}}
  (\bibinfo{publisher}{Clarendon press}, \bibinfo{address}{Oxford},
  \bibinfo{year}{1982}).

\bibitem[{\citenamefont{Brown and Rae}(2014)}]{Brown2014}
\bibinfo{author}{\bibfnamefont{B.}~\bibnamefont{Brown}} \bibnamefont{and}
  \bibinfo{author}{\bibfnamefont{W.}~\bibnamefont{Rae}},
  \bibinfo{journal}{Nuclear Data Sheets} \textbf{\bibinfo{volume}{120}},
  \bibinfo{pages}{115} (\bibinfo{year}{2014}),
  \urlprefix\url{https://doi.org/10.1016/j.nds.2014.07.022}.

\bibitem[{\citenamefont{Honma et~al.}(2009)\citenamefont{Honma, Otsuka,
  Mizusaki, and Hjorth-Jensen}}]{Honma2009}
\bibinfo{author}{\bibfnamefont{M.}~\bibnamefont{Honma}},
  \bibinfo{author}{\bibfnamefont{T.}~\bibnamefont{Otsuka}},
  \bibinfo{author}{\bibfnamefont{T.}~\bibnamefont{Mizusaki}}, \bibnamefont{and}
  \bibinfo{author}{\bibfnamefont{M.}~\bibnamefont{Hjorth-Jensen}},
  \bibinfo{journal}{Physical Review C} \textbf{\bibinfo{volume}{80}}
  (\bibinfo{year}{2009}),
  \urlprefix\url{https://doi.org/10.1103/physrevc.80.064323}.

\bibitem[{\citenamefont{Mukhopadhyay et~al.}(2017)\citenamefont{Mukhopadhyay,
  Crider, Brown, Ashley, Chakraborty, Kumar, McEllistrem, Peters,
  Prados-Est{\'{e}}vez, and Yates}}]{Mukhopadhyay2017}
\bibinfo{author}{\bibfnamefont{S.}~\bibnamefont{Mukhopadhyay}},
  \bibinfo{author}{\bibfnamefont{B.~P.} \bibnamefont{Crider}},
  \bibinfo{author}{\bibfnamefont{B.~A.} \bibnamefont{Brown}},
  \bibinfo{author}{\bibfnamefont{S.~F.} \bibnamefont{Ashley}},
  \bibinfo{author}{\bibfnamefont{A.}~\bibnamefont{Chakraborty}},
  \bibinfo{author}{\bibfnamefont{A.}~\bibnamefont{Kumar}},
  \bibinfo{author}{\bibfnamefont{M.~T.} \bibnamefont{McEllistrem}},
  \bibinfo{author}{\bibfnamefont{E.~E.} \bibnamefont{Peters}},
  \bibinfo{author}{\bibfnamefont{F.~M.} \bibnamefont{Prados-Est{\'{e}}vez}},
  \bibnamefont{and} \bibinfo{author}{\bibfnamefont{S.~W.} \bibnamefont{Yates}},
  \bibinfo{journal}{Physical Review C} \textbf{\bibinfo{volume}{95}}
  (\bibinfo{year}{2017}),
  \urlprefix\url{https://doi.org/10.1103/physrevc.95.014327}.

\end{thebibliography}

\end{document}